\input harvmac
\def\Title#1#2#3{#3\hfill\break \vskip -0.35in
\rightline{#1}\ifx\answ\bigans\nopagenumbers\pageno0\vskip.2in
\else\pageno1\vskip.2in\fi \centerline{\titlefont #2}\vskip .1in}


\def\R{\hbox{\rm I \kern-5pt R}}

\font\ticp=cmcsc10
\def\ajou#1&#2(#3){\ \sl#1\bf#2\rm(19#3)}
%
%
\lref\griff{R.B.~Griffiths, \ajou J. Stat. Phys. &36 (84) 219.}

\lref\grifflogic{R.B.~Griffiths, \ajou Found. Phys. &23 (93) 1601.}

\lref\omnes{R.~Omn\`es, \ajou J. Stat. Phys. &53 (88) 893.}

\lref\omnesbook{R. Omn\`es, {\it The Interpretation of Quantum
Mechanics}, Princeton University Press, Princeton (1994).}

\lref\gmhsantafe{M.~Gell-Mann and J.B.~Hartle in {\it Complexity, Entropy,
and the Physics of Information, SFI Studies in the Sciences of
Complexity}, Vol.
VIII, W.~Zurek (ed.),  Addison Wesley, Reading (1990).}

\lref\gmhthree{M.~Gell-Mann and J.B.~Hartle in {\it Proceedings of the
NATO Workshop on the Physical Origins of Time Asymmetry, Mazag\'on, Spain,
September 30-October 4, 1991}, J.~Halliwell, J.~P\'erez-Mercader, and
W.~Zurek (eds.), Cambridge University Press, Cambridge (1994); gr-qc/9304023.}

\lref\gmhprd{M.~Gell-Mann and J.B.~Hartle, \ajou Phys. Rev. D
&47 (93) 3345.}

\lref\dowkerkentone{F.~Dowker and A.~Kent, 
\ajou J. Stat. Phys. &82 (96) 1575.}

\lref\dowkerkenttwo{F.~Dowker and A.~Kent, \ajou Phys. Rev. Lett. &75
(95) 3038.}

\lref\kentone{A.~Kent, ``Remarks on Consistent Histories and Bohmian
Mechanics'', to appear in 
{\it Bohmian Mechanics and Quantum Theory: An Appraisal}, 
J.~Cushing, A.~Fine and S.~Goldstein (eds.), Kluwer Academic Press;
quant-ph/9511032.}  

\lref\hartleone{J.B.~Hartle, in {\it Quantum Cosmology and Baby
Universes}, Proceedings of the 1989 Jerusalem Winter School on
Theoretical Physics,
ed.~by S.~Coleman, J.~Hartle, T.~Piran, and S.~Weinberg,
World Scientific, Singapore (1991).}

\lref\hartletwo{J.B.~Hartle, \ajou Phys. Rev. &D44 (91) 3173.}

\lref\hartlethree{J.B.~Hartle, {\it Spacetime Quantum Mechanics and the
Quantum Mechanics of Spacetime} in {\it Gravitation and Quantizations:
Proceedings of the 1992 Les Houches
Summer School}, B.~Julia and J.~Zinn-Justin (eds.), North Holland
Publishing Co, Amsterdam, (1994); gr-qc/9304006.}

\lref\saunders{S.~Saunders, {\it ``The Quantum Block Universe''},
Harvard Dept. of Philosophy preprint (1992);
{\it ``Decoherence, Relative States, and Evolutionary Adaptation''},
Harvard Dept. of Philosophy preprint (1993).}

\lref\mwibook{ B.~DeWitt and R.N.~Graham (eds.), {\it The Many Worlds
Interpretation
of Quantum Mechanics}, Princeton University Press, Princeton (1973).}

\lref\everett{H.~Everett, \ajou Rev. Mod. Phys. &29 (57) 454.}

\lref\bellone{J.S.~Bell, {\it ``Quantum Mechanics for Cosmologists''}, in
{\it ``Quantum Gravity 2''}, eds. C.~Isham, R.~Penrose and D.~Sciama,
Clarendon Press, Oxford (1981); reprinted in
J.S.~Bell, {\it Speakable and Unspeakable
in Quantum Mechanics}, Cambridge University Press, Cambridge (1987).}

\lref\belltwo{J.S.~Bell, {\it The Measurement Theory of Everett and
De Broglie's Pilot Wave}
in {\it Quantum Mechanics, Determinism, Causality and Particles}, M.~Flato
et al. (eds.), D.~Reidel, Dordrecht, (1976); reprinted in
J.S.~Bell, {\it Speakable and Unspeakable
in Quantum Mechanics}, Cambridge University Press, Cambridge (1987).}

\lref\stein{H.~Stein, \ajou No\^{u}s &18 (84) 635.}

\lref\ak{A.~Kent, \ajou Int. J. Mod. Phys. &A5 (90) 1745.}

\lref\bohm{D.~Bohm, \ajou Phys. Rev &85 (52) 166.}

\lref\grw{G.~Ghirardi, A.~Rimini and T.~Weber, \ajou Phys. Rev. &D34
(86) 470.}

\lref\hartleone{J.B.~Hartle, in {\it Quantum Cosmology and Baby
Universes}, Proceedings of the 1989 Jerusalem Winter School on
Theoretical Physics,
ed.~by S.~Coleman, J.~Hartle, T.~Piran, and S.~Weinberg,
World Scientific, Singapore (1991).}

\lref\mcelwaine{J.~McElwaine, \ajou Phys. Rev. A &53 (96) 2021.} 


\Title{\vbox{\baselineskip12pt\hbox{ DAMTP/95-65}\hbox{ gr-qc/9512023}{}
}}
{\vbox{\centerline {
Quasiclassical Dynamics in a Closed Quantum System}}}{~}

\centerline{{\ticp Adrian Kent}}
\vskip.1in
\centerline{\sl Department of Applied Mathematics and
Theoretical Physics,}
\centerline{\sl University of Cambridge,}
\centerline{\sl Silver Street, Cambridge CB3 9EW, U.K.}

\bigskip

\centerline{\bf Abstract}
{
We consider Gell-Mann and Hartle's consistent histories 
formulation of quantum cosmology in the  
interpretation in which one history, chosen randomly according
to the decoherence functional probabilities, is realised 
from each consistent set.  
We show that in this interpretation, if one assumes that an observed
quasiclassical structure will continue to be quasiclassical,
one cannot infer that it will obey
the predictions of classical or Copenhagen quantum mechanics. 
}
\newsec{Introduction} 

Modern cosmological theory strongly suggests that large-scale 
classical structure now dominating the universe 
evolved from a highly homogenous quantum state lacking any such structure.  
Since this process cannot be described
in the Copenhagen interpretation of quantum theory, 
other interpretational ideas --- requiring, or claiming to require, no 
pre-existing classical realm --- have attracted increasing attention
over the last forty years.  
There has been particular interest lately in the 
consistent histories approach to quantum theory developed by 
Griffiths,\refs{\griff,\ \grifflogic}
Omn\`es,\refs{\omnes,\ \omnesbook} and 
Gell-Mann and Hartle.\refs{\gmhsantafe,\ \gmhthree}
Unlike earlier suggestive but imprecise proposals in the 
literature, this formulation of the quantum theory of a closed system
admits a well-defined interpretation which defines an interesting 
scientific theory, albeit rather a weak one.  
The purpose of this paper is to explain precisely how weak this theory
is when it comes to predicting the formation and evolution 
of classical structure.  

It was recently shown,\refs{\dowkerkentone} inter alia,
that one cannot use any version of the consistent histories formalism
to predict that the largely classical
structure we observe will persist or appear to persist.  
It is shown here that if we try to evade this difficulty by simply
assuming that 
we will continue to observe a largely classical universe, we cannot 
use the consistent histories formalism to 
predict that the classical equations of motion will hold, even
approximately, or that the results of quantum experiments will 
agree with the predictions of the Copenhagen interpretation. 
The formalism predicts
infinitely many different possible outcomes for a typical classical
or quantum observation or experiment.  The conditional probabilities 
for these outcomes, given the event that classical structure persists
for any fixed time interval, are not defined. 

If the aim is to derive a 
theory of the formation of the observed large-scale structure, or 
its present dynamics, from a consistent histories formulation of
quantum cosmology, these are clearly rather discouraging facts. 
Their implications are discussed in the concluding section.

\newsec{Consistent Histories} 

We simplify the discussion by using
a version of the consistent histories 
formalism in which the initial conditions are defined by a pure
state rather than a density matrix, the basic objects of the
formalism are branch-dependent sets of projections, and consistency
is defined by Gell-Mann and Hartle's decoherence criterion.  
Similar arguments to those below apply in the general 
case.\foot{In particular, if there is an impure
initial density matrix, then the discussion of Ref. \refs{\dowkerkentone}, 
Section 3, can be used to show that a generic quasiclassical 
history belongs to an infinite family of inequivalent consistent 
sets.  The use of branch-dependent sets, though convenient for 
discussing structure formation, is also inessential.}  
The notation is Gell-Mann and Hartle's.\refs{\gmhprd} 

Let $\psi$ be the initial state of the universe.  
A {\it branch-dependent set of histories} is a set of products of
projection operators chosen from projective decompositions and
with a time label.  The set is indexed by the variable 
$\alpha = \{ \alpha_n , \alpha_{n-1} , \ldots , \alpha_1 \}$, where
the ranges of the $\alpha_k$ and the projections they define
depend on the values of $\alpha_{k-1} , 
\ldots , \alpha_1 $, and the histories take the form: 
\eqn\histories{
C_{\alpha} \, = \, P_{\alpha_n}^n (t_n ; \alpha_{n-1} , \ldots , \alpha_1 ) \, 
P_{\alpha_{n-1}}^{n-1} (t_{n-1} ; \alpha_{n-2} , \ldots , \alpha_1 ) \, 
           \ldots \, 
          P_{\alpha_1}^1 ( t_1 )   \, .}
Here, for fixed values of $\alpha_{k-1} , \ldots , \alpha_1$, 
the $P^k_{\alpha_k} \, (t_k ; \alpha_{k-1} , \ldots , \alpha_1 )$ 
define a projective decomposition indexed by $\alpha_k$, so that
$\sum_{\alpha_k} \, P^k_{\alpha_k}  (t_k ; \alpha_{k-1} , \ldots , \alpha_1 )
\, = \, 1 $ and 
\eqn\decomp{
P^k_{\alpha_k}  (t_k ; \alpha_{k-1} , \ldots , \alpha_1 ) \, 
P^k_{\alpha'_k} (t_k ; \alpha_{k-1} , \ldots , \alpha_1 ) \, 
= \, \delta_{\alpha_k  \alpha'_k } \, 
P^k_{\alpha_k} (t_k ; \alpha_{k-1} , \ldots , \alpha_1 ) \, .}
The set of histories is {\it consistent} if and only if 
\eqn\consist{
 ( \, C_{\beta} \psi  \, , \, C_{\alpha} \psi \, ) \, = \, 
\delta_{\alpha \beta } \,  p ( \alpha )   \, , }
in which case $p(\alpha)$ is interpreted as the probability of the 
history $C_{\alpha}$.\foot{Note that, when we use the compact notation
$C_{\alpha}$ to refer to a history, we intend the individual
projection operators, not just their product, to be part of the 
definition of the history.} 
The histories of non-zero probability in a 
consistent set thus
correspond precisely to the non-zero vectors $C_{\alpha} \psi$. 
Only consistent sets are of physical relevance. 
Although the dynamics are defined purely by the hamiltonian, with no
collapse postulate, each projection in the history can be thought of as 
corresponding to a historical event, taking place at the relevant time.  
If a given history is realised, its events correspond to extra physical
information, neither deducible from the state vector nor influencing it. 

Most projection operators involve rather obscure physical quantities,  
so that it is hard to interpret a general history in familiar language.  
Given some sensible model, with hamiltonian and canonical 
variables specified, one can construct sets of histories 
which describe familiar physics and check that they are 
indeed consistent.  For example, a useful set of histories for describing
the solar system could be defined by projection operators whose 
non-zero eigenspace contains states in which a given planet's centre of 
mass is located in a suitably chosen small volumes of space at the 
relevant times, and one would expect a sensible model 
to show that this is a consistent set and that  
the histories of significant probability are those
agreeing with the trajectories predicted by general relativity.  
More generally, Gell-Mann and Hartle\refs{\gmhsantafe} introduce the notion of 
a {\it quasiclassical domain}: a consistent set which is complete ---
in the sense that it cannot be consistently extended by more 
projective decompositions --- and is defined by projection
operators which involve similar variables at different times and
which satisfy classical equations of motion, to a very good 
approximation, most of the time.  
The notion of a quasiclassical domain seems  
natural, though presently imprecisely defined.  Its 
heuristic definition is motivated by the familiar example of the 
hydrodynamic variables --- densities of chemical species in small volumes of 
space, and similar quantities --- which characterise our own 
quasiclassical domain.  Here the branch-dependence of the formalism
plays an important role, since the precise choice of variables 
(most obviously, the sizes of the small volumes) we use depends 
on earlier historical events.  The formation of our galaxy and 
solar system influences all subsequent local physics; 
even present-day quantum experiments have the potential to do 
so significantly, if we arrange for large macroscopic events
to depend on their results. 
It should be stressed at this point that, according to 
all the developers of the consistent histories approach,
quasiclassicality and related properties are interesting 
notions to study within, not defining features of, the formalism.  
All consistent sets of histories have the same physical status. 

By an interpretation of the consistent histories formalism we
mean a description of physics which uses only basic mathematical
quantities defined in the formalism, such as sets and histories, and 
which respects the democracy among consistent sets. 
The literature contains a variety of such interpretations, 
but essentially these are different ways of saying the same 
thing.
One can, with Griffiths, say that precisely one history from each
consistent set is realised, these histories being chosen according
to the probability distribution defined on their set.
One can, more economically, say that in fact only one consistent 
set is physically relevant, but that we have no theoretical rule 
which identifies this set or its properties.\refs{\dowkerkentone}
Or one can, as Gell-Mann and Hartle do, say that the predictions one 
makes depend on the set one uses ---  though here it must be understood
that for almost all sets these predictions will not correspond to the 
physics one actually observes.  
In each case, though, it is to be understood that we can only observe 
events from one history, and that the formalism supplies no
theoretical criterion characterising the consistent set from which
that history is drawn. 
These forms of words are scientifically equivalent.  When we come to 
predicting the future from historical data, our predictions all take 
the form ``if $S$ turns out to be the relevant consistent set, then 
event $E$ will take place with probability $p$''.  No event can be 
predicted independent of the as yet unknown set $S$, and in fact
any prediction made in a generic consistent set $S$ will be incompatible 
with the predictions made in some other consistent 
set $S'$.\foot{These points are
discussed in detail in Ref. \refs{\dowkerkentone}.  
The reader might also find the shorter summaries
in Refs. \refs{\dowkerkenttwo,\ \kentone} useful.} 
We will use the many-histories language here.  Nature consists of a list
of histories $H(S)$ drawn from each consistent set $S$.  No observer 
can observe events from more than one such history.  The formalism predicts
neither the history $H(S)$ in which we find ourselves nor the set
$S$ to which it belongs.  It supplies only the probabilities for 
the possible $H(S)$ given the unknown set $S$. 

I should like to add here the cautionary remark that there is another
interpretation of the consistent histories formalism 
which is {\it not} equivalent to the many-histories interpretation
considered here.\foot{I am very grateful to Todd Brun for pointing
this out in the course of an illuminating correspondence.} 
This interpretation, which as I understand it
is not advocated by Griffiths, Omn\`es, or Gell-Mann \& 
Hartle (but see Saunders\refs{\saunders}), can be summarised 
by saying that {\it every} consistent history is realised in a 
continuum of copies whose measure is given by the history's 
probability weight.\foot{Though the separate discussion necessary 
for this interpretation is beyond the scope of this paper, it seems 
to me that the interpretation suffers from  
related and equally severe problems.}

\newsec{Quasiclassical Histories in Quantum Cosmology} 

In the absence of a quantum theory of gravity, we work in some fixed 
background spacetime with preferred timelike directions and suppose
that the gravitational interactions of matter can be modelled
by a non-covariant quantum potential.   
This is obviously incorrect, 
but at least shares qualitative features with the type of description that
it is hoped might emerge from a fundamental theory.  

A semi-classical treatment, in which the background
manifold depends on the large-scale structures in
the matter distribution described by the different branches,
would presumably give a better description of our quasiclassical
domain.  But it is hard to see any useful role for a consistency
criterion in a semi-classical theory, and in any case no 
adequate semi-classical theory is available. 
Gell-Mann and Hartle's definition of a quasiclassical domain has
to be understood as a definition which 
applies to real world cosmology only in the 
context of a theory of gravity yet to be 
developed.\foot{A discussion of possible generalisations of the 
formalism to quantum gravity can be found in Ref. \refs{\hartleone}.} 

What can be said about cosmology in our model?  
No consistent set can 
give {\it the} correct account of the evolution of large-scale 
structure, since there is no definitive account of the 
unobserved past in the consistent histories formalism.  
However, cosmologically minded consistent historians envisage 
that the set defining our quasiclassical domain can be extended to a
set --- there may well be many such sets, but let us fix on one and
call it $S_0$ ---  which gives a particularly interesting account, 
running very roughly as follows. 

Some projective decomposition $P_{\alpha_1}^1$ at an early time $t_1$ 
characterises inhomogeneities which mark
the beginning of the formation of structure.  
Further decompositions $\, P_{\alpha_2}^2 , \ldots , P_{\alpha_k}^k \,$, 
which depend on the inhomogeneities already realised, 
describe the development of greater and finer-grained inhomogeneity.
By some later time, say $t_{k+1}$, almost all of the projections
in these decompositions become, to a 
very good approximation, projections onto ranges of eigenvalues 
for hydrodynamic variables.  At this point the histories of non-zero
probability define many distinct branches of the quasiclassical 
domain, each of which corresponds to the formation of significantly different 
large-scale structures.  
This branching process continues through to the present,
through processes such as the
quantum spreading of macroscopic bodies and those of their
interactions with microscopic particles or subsystems that
are subsequently macroscopically amplified.  Each probabilistic
quantum experiment that we perform, for example, defines a new branching. 
There are thus, by now, a very large number of non-zero history
vectors $\, C_{\alpha} \, \psi \,$ corresponding to distinct quasiclassical
branches.  
The quasiclassical domain is filled out, 
between all these branchings, 
by many projective decompositions describing
events which are very nearly predictable from the earlier history.

The branching process must stop at some point if the Hilbert space 
is finite-dimensional, since all the non-zero history vectors 
are orthogonal and any new branching adds to their 
number.\refs{\dowkerkentone,\ \dowkerkenttwo} 
The familiar description of quantum experiments 
cannot be reproduced beyond this point, 
since all subsequent events are predictable. 
Indeed, it is not clear that the quasiclassical domain can continue at all. 
Consistent historians thus generally tacitly assume
that the Hilbert space is infinite-dimensional, or at least that the
present number of branches is very much smaller than its dimension.
In order to simplify the discussion we will do so too.

Although Gell-Mann and Hartle generally refer to quasiclassicality
as a property of domains, it is obviously sensible and useful to 
refer to individual histories as being quasiclassical if they are 
built from projectors defined by hydrodynamic variables and 
if the conditional probabilities of most of these projectors, given
the earlier history, is very close to one. 
The picture $\, S_0 \, $ gives, then, is of a large number
of histories $\, C_{\alpha } \, $, including our own 
history $\, C_{\alpha_0} \,$, defined up to the present time $\, t_0 \,$,  
almost all of which are quasiclassical
in their later stages.  
Any quantum experiments we now undertake 
can be described by a consistent set $\, S'_0 \,$ which extends $\, S_0 \,$ by 
projections, defined for the branch of $\, S_0 \,$ 
corresponding to our own history,
which (very nearly) describe future hydrodynamic variables --- the
local densities around the possible paths of a pointer, say ---  that 
record the results.  
The possible outcomes of these
experiments are described by a series of non-zero history vectors 
$\, P_{\beta_1} \, C_{\alpha_0} \psi , \ldots , P_{\beta_k} \, 
C_{\alpha_0} \psi \,$. 
Each of these outcomes corresponds to a quasiclassical
history, which we take to be complete up to time $\, t > t_0 \,$. 
Let us suppose we are just about to undertake such an experiment,
and for simplicity suppose that the number of outcomes, $k$, is 
three or larger. 

Now consider a similar branching, corresponding to another quantum 
process with several 
macroscopically distinct outcomes, described by 
vectors 
$\, P_{\gamma_1} \, C_{\alpha} \psi , \ldots , P_{\gamma_l} \, 
C_{\alpha} \psi \,$
in a history $\, C_{\alpha} \psi \,$ other than our own. 
These histories can be described in the equivalent consistent 
set in which the projective decomposition defined by the
$\, P_{\gamma_i} \,$ is replaced by that defined by the one-dimensional
projectors $\, P'_{\gamma_i} \,$ onto the 
states $\, P_{\gamma_i} \, C_{\alpha}
\psi \,$, together with their 
complement $\, ( 1 \, - \, \sum_i \,  P'_{\gamma_i } ) \,$,
which defines the zero probability  
history  $\, ( 1 \, - \, \sum_i \,  P'_{\gamma_i } ) \, C_{\alpha}
\psi$.  Although there are $(l+1)$ projectors in this decomposition,
there are still only $l$ physical branches, since
zero probability histories are physically irrelevant in
the formalism.  
We can define other consistent sets, which are inequivalent to $\, S_0 \,$ 
and involve non-quasiclassical histories in the branches extending the
history $\, C_{\alpha} \psi \,$, by replacing the 
projectors $\, P'_{\gamma_i} \,$
in this last decomposition by projectors onto any
$l$ states forming an orthogonal basis for the subspace spanned
by the vectors $\, P_{\gamma_i} \, C_{\alpha} \psi \,$. 

By making similar substitutions of the projective decompositions
on all branches other than
our own, we
can construct an infinite number 
of consistent sets $S$ whose only quasiclassical history
is our own, $ \, C_{\alpha_0} \,$. 
After the branching defined, in any of these sets, by the
experiment we are about to undertake, the only quasiclassical
histories will be the $ P_{\beta_i} \, C_{\alpha_0} \,$, corresponding
to the $k$ possible experimental results.  
Finally, we can pick one result, $i_0$, and again define new
consistent 
sets by replacing the projectors $P_{\beta_i}$ for $i \neq i_0$ by
projectors onto another orthonormal basis of the subspace spanned
by $\{ P_{\beta_i }  \, C_{\alpha_0} \, : \, i \neq i_0 \}$. 
In this way we can construct an infinite number of consistent sets 
$\, S \, $ which include precisely one of the 
histories $\, P_{\beta_{i_0}} \, C_{\alpha_0} \,$, which 
extend the history $\, C_{\alpha_0} \,$ non-quasiclassically between 
times $\, t_0 \,$ and $\, t \,$ in all the other $(k-1)$ branches,
and which have no other quasiclassical histories.  
These sets do not correspond to quasiclassical
domains, but contain one quasiclassical branch, which describes
our history to date together with one of the possible
outcomes of the experiments
we are about to 
undertake.\foot{Though sets of histories of this type do not seem to
have been explicitly considered in the literature, most consistent
historians would, I believe, take their existence for granted in
any sensible cosmological model.}

The probabilities of the histories 
$\, P_{\beta_i} \, C_{\alpha_0} \,$ are non-zero.  
Now, according to the many-histories interpretation, one 
history is realised from each of the sets
$\, S \,$, and the probability of being realised from
any given set is simply the standard history probability. 
Since we have seen that there are an infinite number of consistent sets 
containing any of these histories, it follows with probability one 
that each of these $\, k \,$ histories is realised 
infinitely many times from sets $\, S \,$ of the form described above.  
That is, each of the quasiclassical
histories defined by all of our observed data to date, together 
with one of the possible experimental results, is realised infinitely
many times.  The consistent histories formalism, in the many-histories
interpretation, realises an infinite number of copies of each
possible quasiclassical outcome, and these copies of course include
descriptions of ourselves observing our history and the outcome
of the experiment.

This is the problem. 
The formalism supplies us neither with any way of identifying
the correct set from which to draw our history, nor with
any probability measure on the sets.  
Thus, though we can identify the history describing the data we observe, 
and though, when given a particular consistent set, we can calculate 
the probabilities of its histories, we have no way to compute theoretically,
or from empirical data, the probability
of belonging to a history realised from any given set or class of sets.   
If we merely adopt the assumption that our realised history up to 
time $\, t \,$ will be quasiclassical, we can make no 
probabilistic predictions.
In order to do so, we need to make a stronger assumption --- for example,
that our history will be one realised from a particular set. 
To make such an assumption is to go beyond the formalism.

The discussion applies {\it a fortiori} to the predictions of 
classical mechanics since, of course, these predictions are never 
made with complete certainty.
The argument just outlined
holds so long as the probabilities are non-zero, and 
there is always some tiny probability
that the position of a macroscopic object will undergo a
significant quantum fluctuation without violating the 
quasiclassicality of its history. 
While the classical equations of motion are supposed
to hold to a very good approximation, nearly all of the time,
in a quasiclassical domain, a macroscopic
tunnelling event need not violate these criteria. 
For example, if we study a ball thrown against a wall, a
very nearly consistent\foot{
Gell-Mann and Hartle require only approximately consistent 
sets.\refs{\gmhsantafe} 
However, it is conjectured\refs{\dowkerkentone} that this set can
be approximated by an exactly consistent set which describes 
essentially the same physics.  The conjecture is investigated 
further in Ref. \refs{\mcelwaine}.} 
set can be defined by projection operators
whose eigenspaces correspond to states in which the ball's centre
of mass lies within small volumes of space on either side of the
wall, and the history in which the ball's centre of mass 
trajectory goes towards the wall and then continues on the far
side is a quasiclassical history whose probability, though tiny,
is non-zero.  

It might possibly be argued against this last point that it is
sensible to ignore very small probability histories in the 
formalism.  There are several problems with this line of defence,
however.  It is true that, as Gell-Mann and Hartle 
point out,\refs{\gmhsantafe} 
it is often sensible and convenient to ignore small probability 
histories.  If, for example, we have found a good theoretical reason
to fix a particular set of histories for making predictions, and 
find that within that set
the probability of the sun failing to rise tomorrow is
$10^{-10^{40}}$, we can for all practical purposes take it to be zero.
But this does not imply (and Gell-Mann and Hartle do not argue)
that small probability histories are meaningless or always 
theoretically negligible.  In particular, small probability histories
can still give rise to large conditional probabilities. 
The construction above produces infinitely many consistent sets in 
which the only quasiclassical history is one in which the sun fails
to rise and of course, in those sets, the probability of no 
sunrise conditioned on persisting 
quasiclassicality is, tautologically, one.  

It is true that we could simply declare by fiat that all histories with
probability smaller than some parameter $\epsilon$ are to be
neglected.  Some care would be required here, since the probability of our
own realised history, $\epsilon_0$, is by now extremely small, and 
$\epsilon / \epsilon_0 $ would also have to be very small if we are  
to continue observing random events for very long.  But in any case
this strategy would mean that 
$\epsilon / \epsilon_0$ becomes a key parameter in determining the
outcome of experiments.  No outcome $i$ whose probability conditioned
on the past history, $p(i | H )$, is smaller than $\epsilon / \epsilon_0$ 
could arise.  However, we would still have no way of deriving from the 
many-histories interpretation the correct probabilities, 
conditioned on future quasiclassicality, for 
outcomes for which $p(i | H ) > \epsilon / \epsilon_0$. 
Many predictions of classical mechanics might, for some 
finite time interval, be recovered by this strategy --- at the 
price of introducing a new experimentally determinable 
parameter --- but the predictions of Copenhagen quantum mechanics 
cannot be. 

Finally, it should be stressed
that the problem identified here is quite different
from the generally recognised problems of precisely defining
quasiclassicality\refs{\gmhsantafe} and of understanding the 
error limits within which classical physics can be recovered 
once a set involving classical variables has been 
specified.\refs{\omnes, \omnesbook}  The arguments here require 
only a heuristic definition of quasiclassicality,\refs{\gmhsantafe}
but of course would remain valid if a precise definition were supplied.

\newsec{Conclusions} 

The argument we have given is very simple.  
Predictions within the formalism depend on one's choice of set. 
If we choose one of the infinitely many sets whose only 
quasiclassical history describes a series of $\, N \,$ measurements of
$\, \sigma_x \,$ on spin-$1$ particles prepared in the eigenstate 
$\, \sigma_y \, = \, 1 \,$, in each of 
which $\, \sigma_x \, = \, 1 \,$ is observed, 
then our prediction is that either quasiclassicality will fail to
persist or that $\, \sigma_x \, = \, 1 \,$ will repeatedly be observed. 
If we condition on the persistence of quasiclassicality, then in
this set the latter prediction is made with probability one.
And indeed, this sequence of results is realised 
in an infinite number of sets, as are all other sequences.  
Without a measure on the space of sets, we cannot assign any
{\it a priori} probability distribution to the choice of set
which should be used for prediction, and hence --- if we 
assume the persistence of quasiclassicality --- cannot 
assign any probabilities to our quasiclassical predictions.

Is the conclusion interesting?  Why should anyone have hoped to
calculate conditional probabilities of the type considered? 
Might the conclusion
perhaps rely on a perverse reading of the consistent
histories formalism?  Can we not easily find
another interpretation in which no similar difficulty arises?
Is there perhaps a natural measure on 
the consistent sets which produces the correct 
probabilistic predictions?
If not, is there a simple amendment to the formalism which
does the job?  
We take these points in turn. 

At issue here is the relation between
the consistent histories formulation of quantum cosmology, classical
mechanics and Copenhagen quantum mechanics.  Nothing that the 
consistent histories formalism
says is inconsistent with either of these last two theories.
It does not contradict their predictions.  However, it does
not allow us to derive them.  Given any quasiclassical history, such
as the one we find ourselves in, the formalism makes no probabilistic
or deterministic predictions of future events.  As we have seen,
this still holds true if we assume that the history will continue to 
be quasiclassical.  The predictions of Copenhagen quantum mechanics
do not follow even from the consistent histories account of quantum
cosmology combined with the assumption of a quasiclassical history
obeying standard classical mechanics to a good approximation.  
All three theories are independent.  

Gell-Mann and Hartle argue\refs{\gmhsantafe} that, although all 
consistent sets are equivalent in the formalism, we find ourselves 
perceiving a quasiclassical history because we have evolved so as
to become sensitive to quasiclassical variables and adapted to make
use of them.   There are implicit assumptions in this 
argument which
need not concern us here.\refs{\dowkerkentone}  
Let us accept that it might be so, and 
suppose that some theory of perception tells us that for the 
purpose of predicting our own future perceptions we can ignore the
possibility that we might find ourselves in a non-quasiclassical history.
The preceding discussion still tells us that there are
infinitely many observers sharing our evolutionary history,
continuing to observe a quasiclassical world in the future, 
who find their subsequent observations disagreeing with classical 
mechanics and Copenhagen quantum mechanics.  This in itself need not be 
an insuperable problem; however, the formalism does not define
any probability measure that allows us to tell which type of 
realised quasiclassical history is more probable.  
Thus, accepting Gell-Mann and Hartle's 
argument, we find ourselves unable to use the consistent histories
formalism to make the predictions of classical mechanics
and Copenhagen quantum mechanics. 

We need not interpret the formalism in many-histories language.
The other interpretations of the formalism
in the literature, though, have the same
implication when it comes to making predictions, and
it is easy to see why without 
rehearsing the full argument\refs{\dowkerkentone} for their equivalence.
To predict anything, in any interpretation, we require data, in the form of 
the observed history $\, H \,$, and a consistent set $\, S \,$ which 
includes the
projections defining that history.  Neither the inclusion of the 
projections defining $\, H \,$, nor the assumption that $\, S \,$ contains 
quasiclassical histories extending $\, H \,$, are very 
strong constraints on $\, S \,$.
Without assuming some sort of probability measure on the space of sets
we cannot characterise the likely properties of $\, S \,$.  In particular
we cannot say what type of histories it is likely to contain, or which
quasiclassical histories it is likely to contain. 

There is, indeed, a natural measure on the consistent sets of histories,
defined at least if the Hilbert space is finite-dimensional, and inherited
from the geometric description of consistent sets as algebraic 
curves.\refs{\dowkerkentone}  
Unfortunately, though unsurprisingly, when restricted 
to the class of sets containing extensions of a given quasiclassical
history, it assigns measure zero to the subclass containing
quasiclassical extensions of that history.  In other words, if the
formalism is amended so that the physically relevant set is chosen 
according to the natural measure, it predicts with probability one
that the quasiclassicality we observe will cease immediately.  

The obvious amendment to the formalism is to abandon democracy
amongst consistent sets.
If we hypothesise that a set defining our
quasiclassical domain is {\it the} physically relevant set, or 
more generally that among the physically relevant sets it is the
only one including some the projections which characterise the observed
data, then we can certainly predict the persistence of 
quasiclassicality and derive the predictions of classical and Copenhagen
quantum mechanics.\foot{Another possible strategy, though one fraught
with problems, is the selection of sets which characterise observers
rather than domains.\refs{\dowkerkentone}}  
In practice this is almost precisely what we do when we make
experimentally testable 
predictions: we do not typically use all the projections defining
our quasiclassical domain, but the variables we do consider are
always quasiclassical projections or operators almost perfectly
correlated with those projections. 
To suppose that a particular set or type of set is fundamentally
preferred, of course, is to go beyond orthodox quantum
theory, by insisting that particular variables are distinguished.  
However, it appears to be necessary in order
to derive our most successful physical theories from 
the consistent histories formulation of quantum cosmology. 

There are at least two genuine, and genuinely new, 
interpretations of 
quantum theory which follow the line of
thought that begins with the arguments of Everett 
et al.\refs{\mwibook} 
that quantum theory admits a ``many-worlds interpretation''. 
One of these, due to Bell,\refs{\bellone,\ \belltwo} abandons the notion of 
a coherent historical description of physics entirely: the events
occurring at any time are uncorrelated with those at earlier or
later times.  This proposal is logically consistent and, given 
the correct dynamics and boundary conditions, experimentally
unfalsifiable, but is not thought by most physicists (and was not
thought by Bell) to be a useful scientific theory, since it 
makes cosmology redundant, memory fictitious, and useful
prediction impossible.  
The other is the interpretation based on the
consistent histories formalism considered here.  Neither allows 
the derivation of a theory of quasiclassical physics.  

This is not to say that either
the formalism itself or the current ideas about structure formation 
are misguided.  The former 
suggests at least one possible way of going beyond orthodox quantum 
theory.  
The latter implicitly rely on intuitions, which may well be 
sound, about the variables which might be distinguished. 
It would seem, though, that if we want a genuine derivation
of a theory of the formation and dynamics of the quasiclassical structure
in the universe from quantum cosmology, in which we can make the usual
quasiclassical predictions, then we have to go beyond orthodox
quantum theory as it is presently understood by identifying preferred
variables in some way.  This need not necessarily 
involve any change in the dynamics. 

\vskip.3in

\leftline{\bf Acknowledgements}  

I am very grateful to Fay Dowker for many helpful discussions 
and to Todd Brun, Bob Griffiths, Jim McElwaine, Trevor Samols and anonymous
referees for valuable comments.  
This work was supported by a Royal Society University 
Research Fellowship. 
\listrefs
\end